\documentclass[twocolumn,showpacs,preprintnumbers,amsmath,amssymb,epsf]{revtex4}
\usepackage{graphicx}
\usepackage{dcolumn}
\usepackage{bm}
\begin{document}

\title{Doubly charged Higgs through photon-photon collisions in 3-3-1 model}
\author{J. E. Cieza Montalvo$^1$, Nelson V. Cortez$^2$, M. D. Tonasse$^{3}$}
\address{$^1$Instituto de F\'{\i}sica, Universidade do Estado do Rio de Janeiro, Rua S\~ao Francisco Xavier 524, 20559-900 Rio de Janeiro, RJ, Brazil}
\address{$^2$Justino Boschetti 40, 02205-050 S\~ao Paulo, SP, Brazil}
\address{$^3$Unidade de Registro, {\it Campus} Experimental de Registro, Universidade Estadual Paulista, Rua Tamekishi Takano 5, 11900-000, Registro, SP, Brazil}
\date{\today}
\begin{abstract}
We study the production and signatures of doubly charged Higgs bosons in the process $\gamma \gamma \leftrightarrow H^{--} H^{++}$ at the $e^-e^+$ 
International Linear Collider and CERN Linear Collider,  where the intermediate photons are given by the Weizs$\ddot{a}$cker-Willians and laser backscattering distributions. 
\end{abstract}
\maketitle
\section{INTRODUCTION \label{introd}}
The Higgs boson is the one missing piece, is a critical ingredient to complete our understanding of the Standard Model (SM), the current theory of fundamental particles and how they interact. Different types of Higgs bosons, if they exist, may lead us into new realms of physics beyond the SM. Up to now, SM experiments have given no evidence for the presence of the SM  Higgs boson. So, the observation of any kind of Higgs particle must be an important step forward in the understanding of physics in the electroweak sector or beyond the SM. In extended models doubly charged Higgs bosons (DCHBs) appear as being relatively light and  typically as a component of triplet representations such as the Left-Right  Symmetric Models \cite{pati}. They appear also as components of sextet ones such as SU(3)$_L \otimes$ U(1)$_N$ (3-3-1) \cite{PP92}. It  was shown that the simplest 3-3-1 model, which is also the simplest chiral extention of the SM, is in triplet representation of $SU(3)$ of the matter field, so it  incorporates the  DCHBs naturally and the results in Refs. \cite{cnt1,cnt2} indicate a sufficient number of events required for establishing the signal. Researches has been made with DCHBs and a lower limit of $141(114)$ GeV for their masses are obtained by Hera and Fermilab respectivelly \cite{aktas}. \par

Independently of DCHBs, the 3-3-1 models have interesting features for the TeV scale energy. Let us briefly enumerate some notable points of them: {\bf a)} Due to cancellation  of chiral anomalies, the number of generations $N$ must be a multiple of 3, however due to asymptotic freedom in QCD, which impose that the number of generations must be $N$ $\le$ 5, the unique allowed number of generations is $N=3$; {\bf b)} Some of the 3-3-1 process accessible to next generation of accelerators violate individual lepton numbers; \ {\bf c)} The quantization of electric charge is independend if neutrinos are Dirac or Majorana particles \cite{ppp}; \ {\bf d)} The seesaw mechanism can be naturally incorporate in some version of the 3-3-1 models \cite{CT05}; \ {\bf e)} The building of the supersymmetric version of a 3-3-1 gauge model \cite{331SM02}; \ {\bf f)} The models are non-perturbative at TeV scale, as well as the supersymmetric version \cite{vicente}; \ {\bf g)} We observe that the parameter $sin^2 \theta_W /(1-4sin^2 \theta_W)$ has a Landau pole, that is, $\sin^2 \theta_W$ must be  $< 1/4$, that is a good feature of the model, since evolving the $\sin^2 \theta_W$ to high values it is possible to find an upper bound to the masses of the new particles \cite{vicente}. \par 

The simplest scalar sector of the 3-3-1 model has three Higgs triplets \cite{PT93a,PT93b}, in this version the three scalar triplets are $\eta = \left(\begin{array}{ccc} \eta^0 & \eta^-_1 & \eta^+_2\end{array}\right)^{\tt T}$, $\rho = \left(\begin{array}{ccc} \rho^+ & \rho^0 & \rho^{++}\end{array}\right)^{\tt T}$ and $\chi = \left(\begin{array}{ccc} \chi^- & \chi^{--} & \chi^0\end{array}\right)^{\tt T}$ transforming as $\left({\bf 3}, 0\right)$, $\left({\bf 3}, 1\right)$ and $\left({\bf 3}, -1\right)$, respectively. The neutral components of the scalars triplets $\eta$, $\rho$ and $\chi$ develop non zero vacuum expectation values $\langle\eta^0\rangle = v_\eta$, $\langle\rho^0\rangle = v_\rho$ and $\langle\chi^0\rangle = v_\chi$, with $v_\eta^2 + v_\rho^2 = v_W^2 = (246 \mbox{ GeV})^2$. The pattern of symmetry breaking is $\mbox{SU(3)}_L \otimes\mbox{U(1)}_N\stackrel{\langle\chi\rangle}{\longmapsto} \mbox{SU(2)}_L\otimes\mbox{U(1)}_Y\stackrel{\langle\eta, \rho\rangle}{\longmapsto}\mbox{U(1)}_{\rm em}$. In the potential \cite{ton1} $f$ is a constant with dimension of mass and the $\lambda_i$ $\left(i = 1, \dots, 9\right)$ are adimensional constants, which represents  the intensities of the quartic vertices of the Higgs, with $\lambda_3 < 0$ and $f < 0$ from the positivity of the scalar masses. \par

In the case of elastic $e^{-} e^{+}$ scattering we will use the $\gamma
\gamma$ differential luminosity which is given by 
\begin{eqnarray}
\left (\frac{d \rm L^{el}}{d\tau} \right )_{\gamma \gamma/\ell \ell}  =  \int_{\tau}^{1} \frac{dx_{1}}{x_{1}} f_{\gamma/\ell} (x_{1})    
f_{\gamma/\ell} (x_{2} = \tau/x_{1})   \; ,
\end{eqnarray}  
where $\tau = x_{1} x_{2}$ and $f_{\gamma/\ell} (x)$ are the distribution functions \cite{bs,lb}. The presentation of this subject is given by \cite{bail}.

Studies of DCHBs using specifically $\gamma\gamma$ collision were also done in Ref. \cite{chak}.\par 

In the present work we report about signals of this kind of bosons, which appear when we consider some particular values for the adimensional parameter $\lambda_9$ in the Higgs potential, the vacuum expectation value $v_\chi$ and some masses of DCHBs. \par


\section{CROSS SECTION PRODUCTION}

We study the production of DCHBs in the process $e^{-} e^{+} \rightarrow e^{-} e^{+} \gamma \gamma \rightarrow H^{\pm \pm} H^{\pm \pm}$, which occurs through the non-resonant contributions of the $H^{\pm \pm}$ and $U^{\pm \pm}$ in the $t$ and $u$ channels and the quartic vertex $\gamma \gamma H^{--} H^{++}$. The intermediate photons are generated either by the Weizs$\ddot{a}$cker-Willians \cite{bs} or laser backscattering distributions \cite{lb}. The interaction Lagrangian is given in several papers (see for example, Ref. \cite{cnt1}), then we evaluate the differential subprocess cross section for this reaction as

\hskip 0.6cm
\begin{widetext}
\begin{eqnarray} 
\frac{d \hat{\sigma}}{d\cos \theta} & = & \frac{2 \beta  \  \alpha^{4} \ \pi^{3} [(\Lambda_{U\gamma})^{\mu \nu} (\Lambda_{U\gamma})_{\mu \nu}]^{2}}{\hat{s}}(\frac{\hat{t^{2}}}{m_{U^{\pm \pm}}^{4}}- \frac{2 \hat{t}}{m_{U^{\pm \pm}}^{2}}+ 4 )\left[\frac{1}{(\hat{t}- m_{U^{\pm \pm}}^{2})^{2}} + \frac{1}{(\hat{u}- m_{U^{\pm \pm}}^{2})^{2}}  \right.  \nonumber  \\ 
&&\left. + \frac{4}{(\hat{t}- m_{U^{\pm \pm}}^{2})(\hat{u}- m_{U^{\pm \pm}}^{2})} \right] +  \frac{\beta  \  \alpha^{2} \ \pi [(\Lambda_{\gamma})^{\mu} (\Lambda_{\gamma})_{\mu}]^{2}}{8\hat{s}}  \left [\frac{\hat{t}^{2}}{(\hat{t}- m_{H^{\pm \pm}}^{2})^{2}} + \frac{\hat{u}^{2}}{(\hat{u}- m_{H^{\pm \pm}}^{2})^{2}}  \right.  \nonumber   \\
&&\left. +\frac{2 m_{H^{\pm \pm}}^{4}}{(\hat{t}- m_{H^{\pm \pm}}^{2})(\hat{u}- m_{H^{\pm \pm}}^{2})}  \right] + \frac{\beta  \  \alpha^{3} \ \pi^{3} (\Lambda_{\gamma})^{\mu} (\Lambda_{\gamma})_{\mu} (\Lambda_{U\gamma})^{\mu \nu} (\Lambda_{U\gamma})_{\mu \nu}}{\hat{s}} \left [(\frac{\hat{t}^{2}}{m_{U^{\pm \pm}}^{2}}- \hat{t})  \right.   \nonumber \\
&&\left. (\frac{1}{(\hat{t}- m_{U^{\pm \pm}}^{2})(\hat{t}- m_{H^{\pm \pm}}^{2})}) + (\frac{m_{H^{\pm \pm}}^{4}}{m_{U^{\pm \pm}}^{2}}- \hat{u})(\frac{1}{(\hat{t}- m_{U^{\pm \pm}}^{2})(\hat{u}- m_{H^{\pm \pm}}^{2})}) + (\frac{\hat{t}^{2}}{m_{U^{\pm \pm}}^{2}}- \hat{t}) \right.  \nonumber  \\
&&\left. (\frac{1}{(\hat{u}- m_{U^{\pm \pm}}^{2})(\hat{t}- m_{H^{\pm \pm}}^{2})}) + (\frac{m_{H^{\pm \pm}}^{4}}{m_{U^{\pm \pm}}^{2}}- \hat{u})(\frac{1}{(\hat{u}- m_{U^{\pm \pm}}^{2})(\hat{u}- m_{H^{\pm \pm}}^{2})})  \right]  \nonumber  \\
&& + \frac {4 \pi \beta \alpha^{2}}{\hat{s}}\left [8 + 4 \pi \alpha (\Lambda_{U\gamma})^{\mu \nu} (\Lambda_{U\gamma})_{\mu \nu} (\frac{\hat{t}}{m_{U^{\pm \pm}}^{2}} -4 )(\frac{1}{\hat{t}- m_{U^{\pm \pm}}^{2}} + \frac{1}{\hat{u}- m_{U^{\pm \pm}}^{2}} ) \right. \nonumber \\
&&\left. - (\Lambda_{\gamma})^{\mu} (\Lambda_{\gamma})_{\mu} (\frac{\hat{t}}{\hat{t}- m_{U^{\pm \pm}}^{2}}  + \frac{\hat{u}}{\hat{u}- m_{U^{\pm \pm}}^{2}}) \right],
\end{eqnarray}
\end{widetext}
where $\alpha$ is the fine structure constant, which we take equal to  $\alpha =1/128$, $\sqrt{\hat{s}}$ is the center of mass energy of the $\gamma \gamma$ system, $\hat{t}= m_{H^{\pm \pm}}^{2} - (1 - \beta \cos \theta)\hat{s}/2$ and $\hat{u} = m_{H^{\pm \pm}}^{2} - (1 + \beta \cos \theta)\hat{s}/2$, with $\beta$ being the velocity of the Higgs in the subprocess c. m. and $\theta$ its angle with  respect to the incident $\gamma$ in this frame, $(\Lambda_{U\gamma})_{\mu \nu}$ is the vertex  strength of bosons $U^{\pm \pm}$ to $\gamma$ and $H^{\pm \pm}$, $(\Lambda_{\gamma})_{\mu}$ is the vertex strength of the $H^{\pm \pm}$ to $\gamma$ and $H^{\pm \pm}$, and the quartic vertex strength is equal to $-4i e^{2} g_{\mu \nu}$. The analytical expressions for these vertex strengths are

\begin{eqnarray}
(\Lambda_{U\gamma})_{\mu \nu} & = & i\frac{v_{\eta} v_{\chi}}{\sin \theta_{W}} \sqrt{\frac{2}{v_{\eta}^{2}+ v_{\chi}^{2}}} \ g_{\mu \nu},    \\ 
(\Lambda_{\gamma})_{\mu} & = &  \frac{v_{\chi}^{2}- v_{\eta}^{2}}{v_{\chi}^{2}+ v_{\eta}^{2}}  \ (p_{1}- q_{1})_{\mu}   \  .
\end{eqnarray}
where $p_{1}$ and $q_{1}$ are the momentum four-vectors of the $\gamma$ and $H^{\pm \pm}$ and the others couplings are given in \cite{cnt1,cieneto}.  
For the standard model parameters we assume PDG values, {\it i. e.},
$\sin^2{\theta_W} = 0.2315$ \cite{Cea98}. We can obtain the total cross section for this process folding $\hat{\sigma}$ with the two photon luminosities 
\begin{eqnarray}
\sigma &=& \int_{\tau_{min}}^{1} \frac{d \it{L}}{d \tau} d\tau \ \hat{\sigma} (\hat{s} x_{1} x_{2} s) \nonumber \\
&=& \int_{\tau_{min}}^{1}  \int_{\ln{\sqrt(\tau)}}^{-\ln{\sqrt(\tau)}} \frac{dx_{1}}{x_{1}} f_{\gamma/\ell} (x_{1}) f_{\gamma/\ell} (x_{2}) . \nonumber \\
&& . \int \frac{d \hat{\sigma}}{d \cos \theta}  \ d \cos \theta 
\end{eqnarray}
\section{RESULTS AND CONCLUSIONS}
In the following we present the cross section for the process $e^{-} e^{+} \rightarrow e^{-} e^{+} \gamma \gamma \rightarrow H^{\pm \pm} H^{\mp \mp}$ for the International Linear Collider (ILC) (1.5 TeV) and CERN Linear Collider (CLIC) (3 TeV), where we have chosen for the parameters, masses and the VEV, the following representative values: $\lambda_{1} =-1.2$,  $\lambda_{2}=\lambda_{3}=-\lambda_{6}=\lambda_{8}=-1$, $\lambda_{4}= 2.98$ $\lambda_{5}=-1.57$, $\lambda_{7}=-2$ and $\lambda_{9}=-1.2$, $v_{\eta}=195$ GeV, $v_{\chi}=1000$ and $v_{\chi}=1500$ GeV and with other particles masses as given in the Table \ref{tab1}, it is to notice that the value of $\lambda_{9}$ was chosen this way in order to guarantee the approximation $-f \simeq v_{\chi}$, \cite{ton1} and that the masses of $m_{h^0}$, $m_{H_1^{\pm}}$ and $m_{H_{2}^{\pm}}$  depend on the parameter $f$ and therefore they can not be fixed by any value of $v_{\chi}$, so when we have $M_{H^{\pm \pm}}= 500$ GeV, $v_{\chi}=1000$ GeV, the masses of $H_{2}^{\pm}$ and $h$ are  $m_{H_{2}}^{\pm}= 671.9$ GeV, and $m_h = 1756.2$ GeV, and in the case of $v_{\chi}=1500$ for $M_{H^{\pm \pm}}= 500$ GeV, the values of the mass of $H_{2}^{\pm}$ and $h$ are $m_{H_{2}}^{\pm}= 901.6$ GeV and $m_h = 2802.7$ GeV, respectively.\par

\begin{widetext}
\begin{center}
\begin{table} [h]
\caption{\label{tab1}\footnotesize\baselineskip = 12pt 
Values for $v_{\chi}$, for the masses of heavy leptons (E,M and T), Higgs bosons ($H_{0}^{2}$ and $H_{0}^{3}$), gauge bosons (V,U and Z') and heavy quarks ($J_{1}$, $J_{2}$ and $J_{3}$) for $v_\eta = 195$ GeV. All the values in this table are given in GeV.}
\begin{tabular}{ccccccccccccc}  
\hline\hline
$v_\chi$ & $m_E$ & $m_M$ & $m_T$ & $m_{H^0_2}$ & $m_{H^0_3}$ & $m_V$ & $m_U$ & $m_{Z^\prime}$ & $m_{J_1}$ & $m_{J_2}$ & $m_{J_3}$\\ 
1000 & 148.9 & 875 & 2000 & 1017.2 & 2000 & 467.5 & 464 & 1707.6 & 1000 & 1410 & 1410 \\
1500 & 223.3 & 1312.5 & 3000 & 1525.8 & 3000 & 694.1 & 691.8 & 2561.3 & 1500 & 2115 & 2115 \\
\hline\hline
\end{tabular}
\end{table}
\end{center}
\end{widetext}

The Figs. (1 to 9) will show the behaviour of the cross section of the process $e^{+} e^{-} \rightarrow e^{-} e^{+} \gamma \gamma  \rightarrow  H^{\pm \pm} H^{\mp \mp}$ as a function of $m_{H^{\pm \pm}}$ for bremstrahlung and laser backscattering photons respectively. In that case, the cross section for the process initiated by backscattered photons is approximately one to two orders of magnitude larger than the one for bremstrahlung photons due to the distribution of backscattered photons being harder than the one for bremstrahlung. So in Fig. 1, we show the cross section for ILC for bremstrahlung distribution. \par

\begin{figure}
\includegraphics [scale=.55]{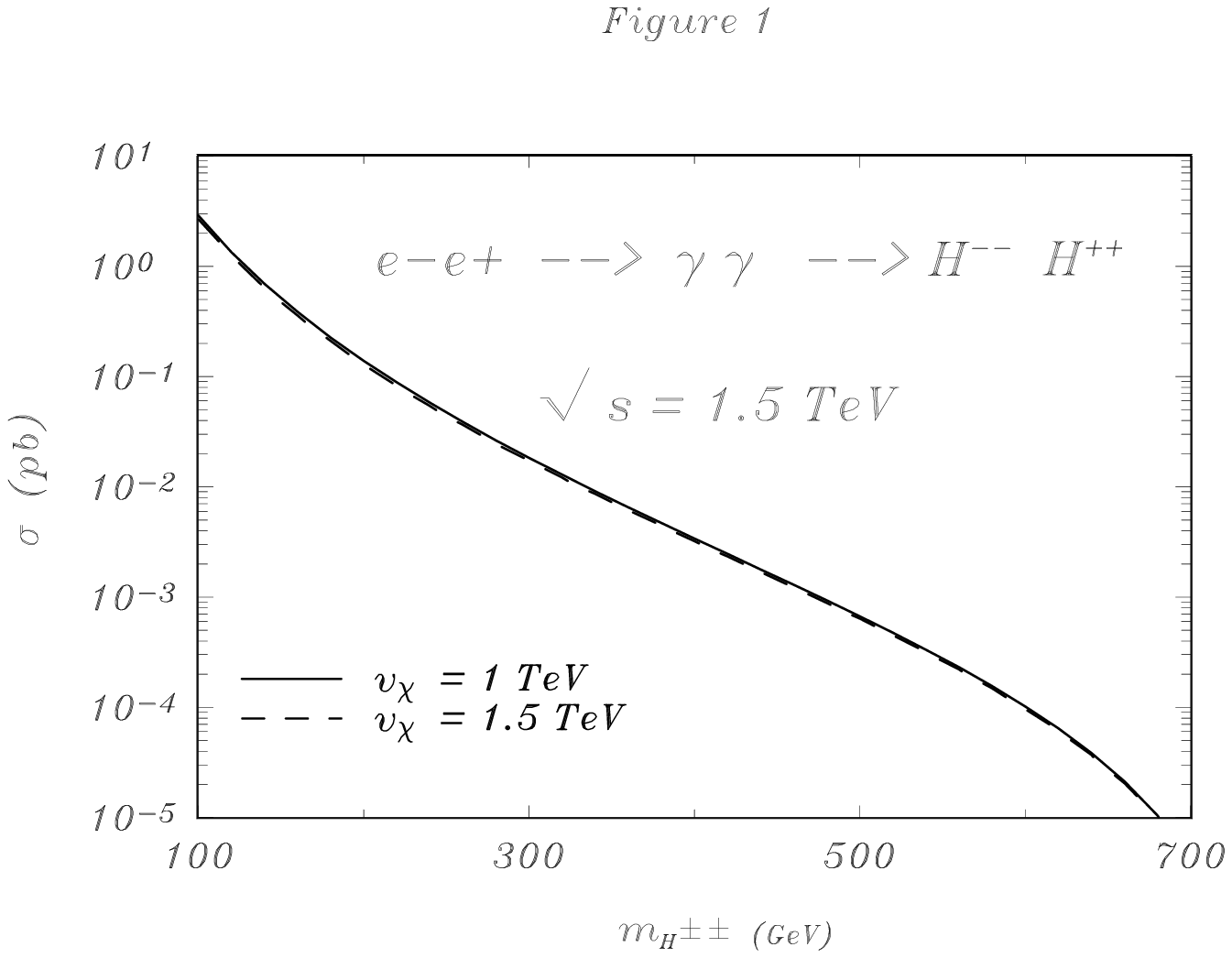}
\caption{\label{fig1} Total cross section for the process $e^{-} e^{+} \rightarrow e^{-} e^{+} \gamma \gamma \rightarrow H^{\pm \pm} H^{\mp \mp}$ as function of $ m_{H^{\pm \pm}}$ for bremstrahlung at $\sqrt{s}=1.5$ TeV a) $v_{\chi}=1$ TeV (solid line) and b) $v_{\chi}=1.5$ TeV (dashed line).}
\end{figure}

\begin{figure}
\includegraphics [scale=.55]{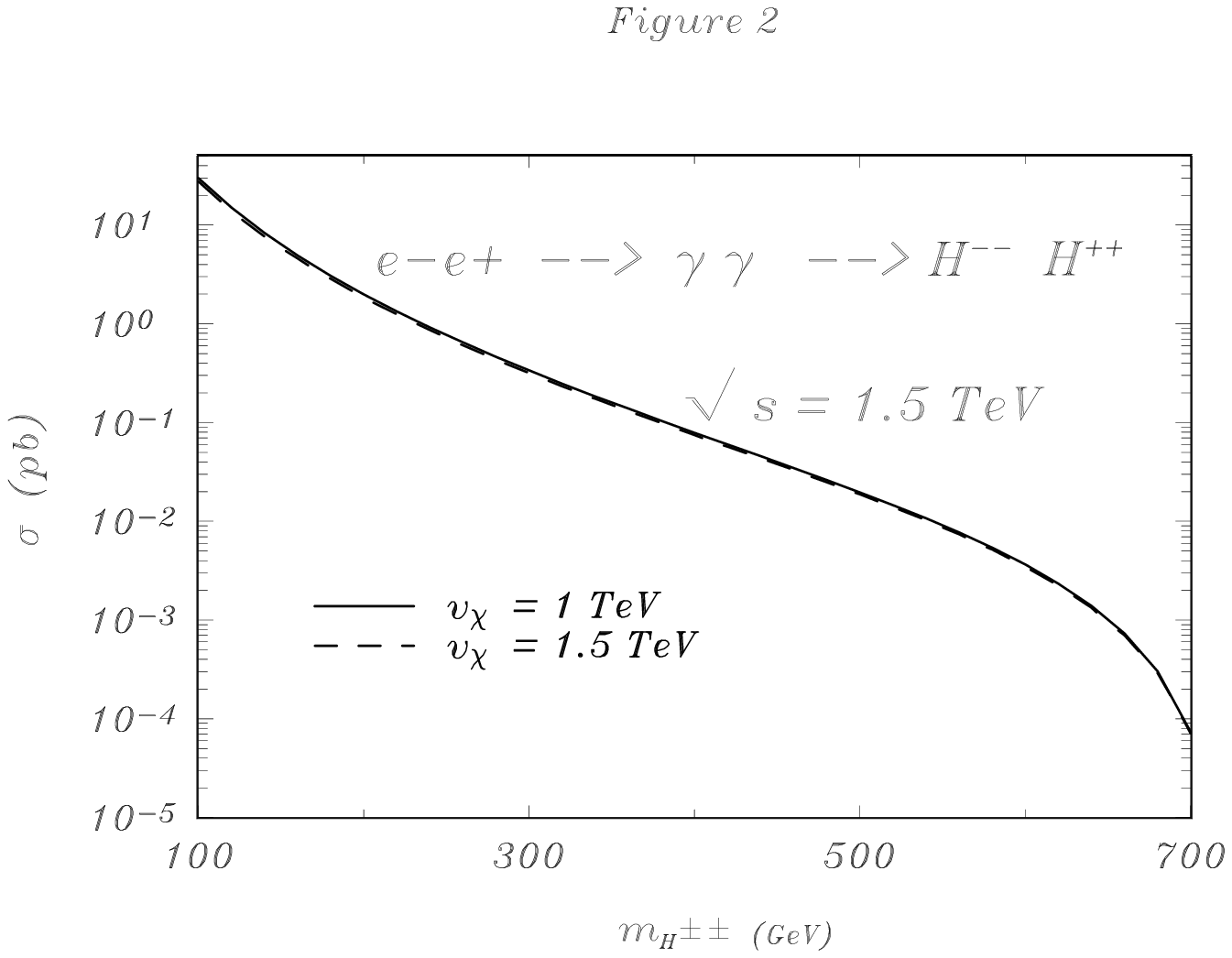}
\caption{\label{fig2} Total cross section for the process $e^{-} e^{+} \rightarrow e^{-} e^{+} \gamma \gamma \rightarrow H^{\pm \pm} H^{\mp \mp}$ as function of $ m_{H^{\pm \pm}}$ for backscattered photons at $\sqrt{s}=1.5$ TeV a) $v_{\chi}=1$ TeV (solid line) and b) $v_{\chi}=1.5$ TeV (dashed line).}
\end{figure}

\begin{figure}
\includegraphics [scale=.55]{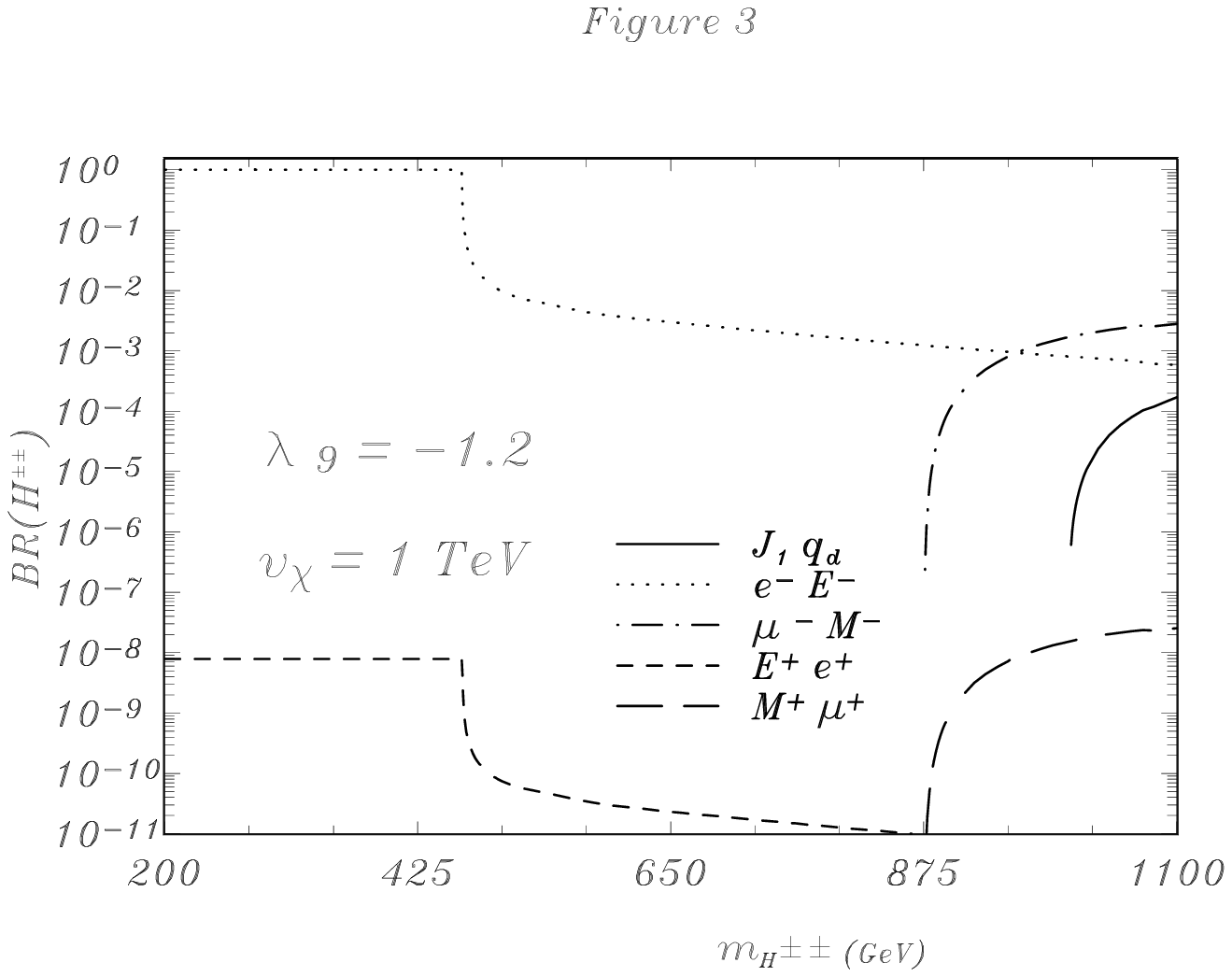}
\caption{\label{fig3} Branching ratios for the doubly charged Higgs decays as functions of $m_{H^{\pm \pm}}$ for $\lambda_{9}=- 1.2$, $v_{\chi}=1$ TeV for the quarks ($J_{1} \ q_{d}$) and leptons ($e^{-} \ E^{-}$, $\mu^{-} \  M^{-}$, $E^{+} \ e^{+}$ and $M^{+} \ \mu^{+}$).}
\end{figure}

\begin{figure}
\includegraphics [scale=.55]{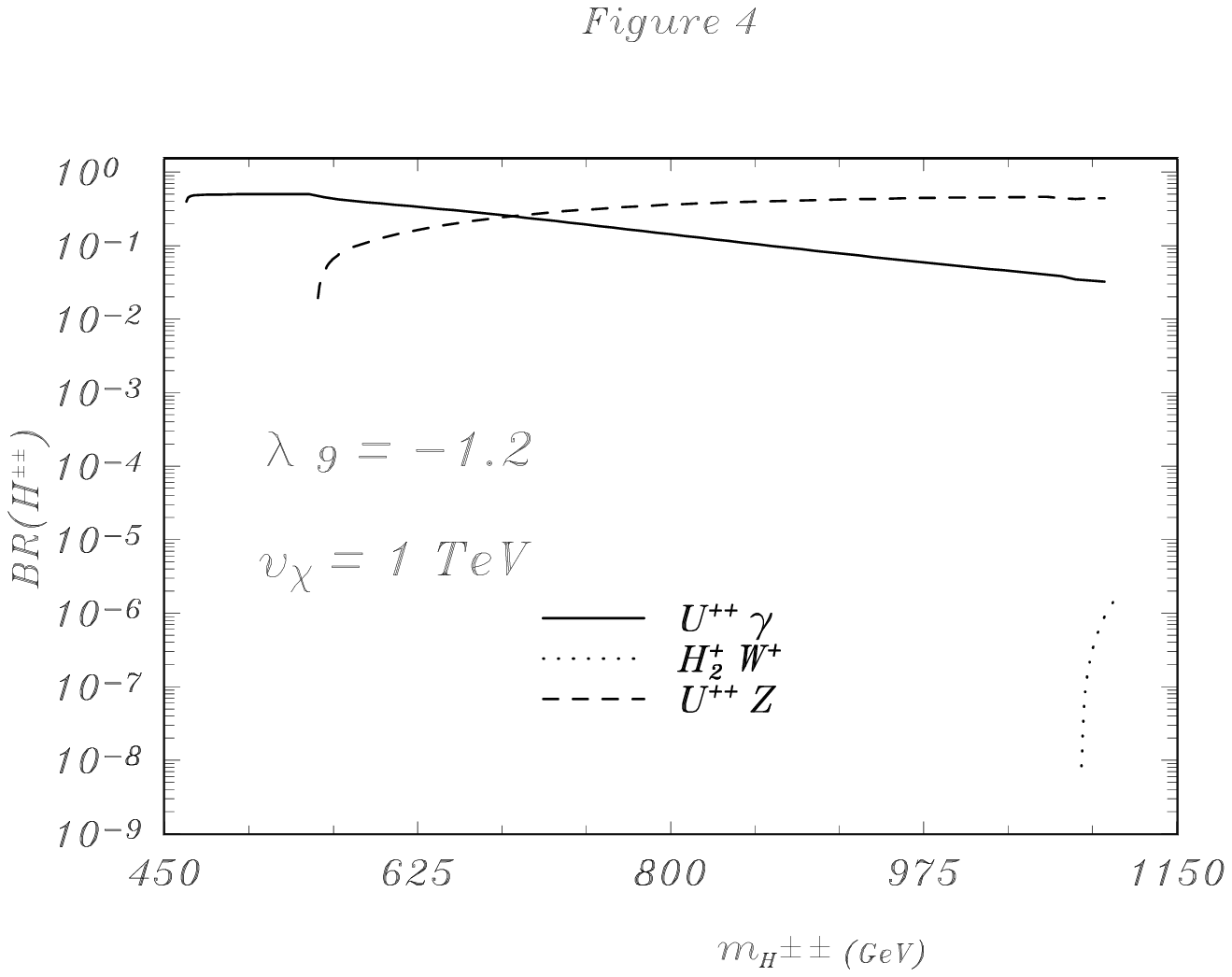}
\caption{\label{fig4} Branching ratios for the doubly charged Higgs decays as functions of $m_{H^{\pm \pm}}$ for $\lambda_{9}=- 1.2$, $v_{\chi}=1$ TeV, for the new gauge bosons and foton,  Z ($U^{\pm \pm} \ \gamma$ and $Z$), for the Higgs boson and gauge boson ($H_{2}^{+} \ W^{+}$).}
\end{figure}

\begin{figure}
\includegraphics [scale=.55]{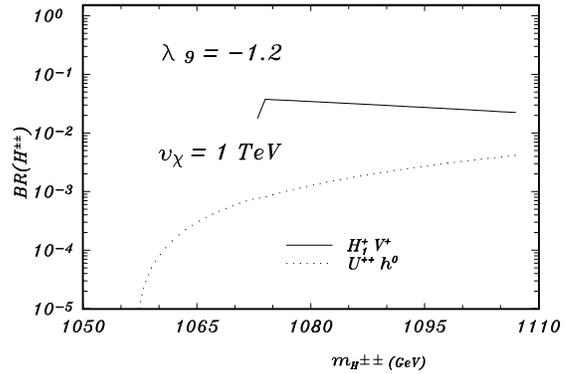}
\caption{\label{fig5} Branching ratios for the doubly charged Higgs decays as functions of $m_{H^{\pm \pm}}$ for $\lambda_{9}=- 1.2$, $v_{\chi}=1$ TeV for $H_{1}^{\pm} V^{\pm}$ and $U^{+ +}h^{0}$.}
\end{figure}

\begin{figure}
\includegraphics [scale=.55]{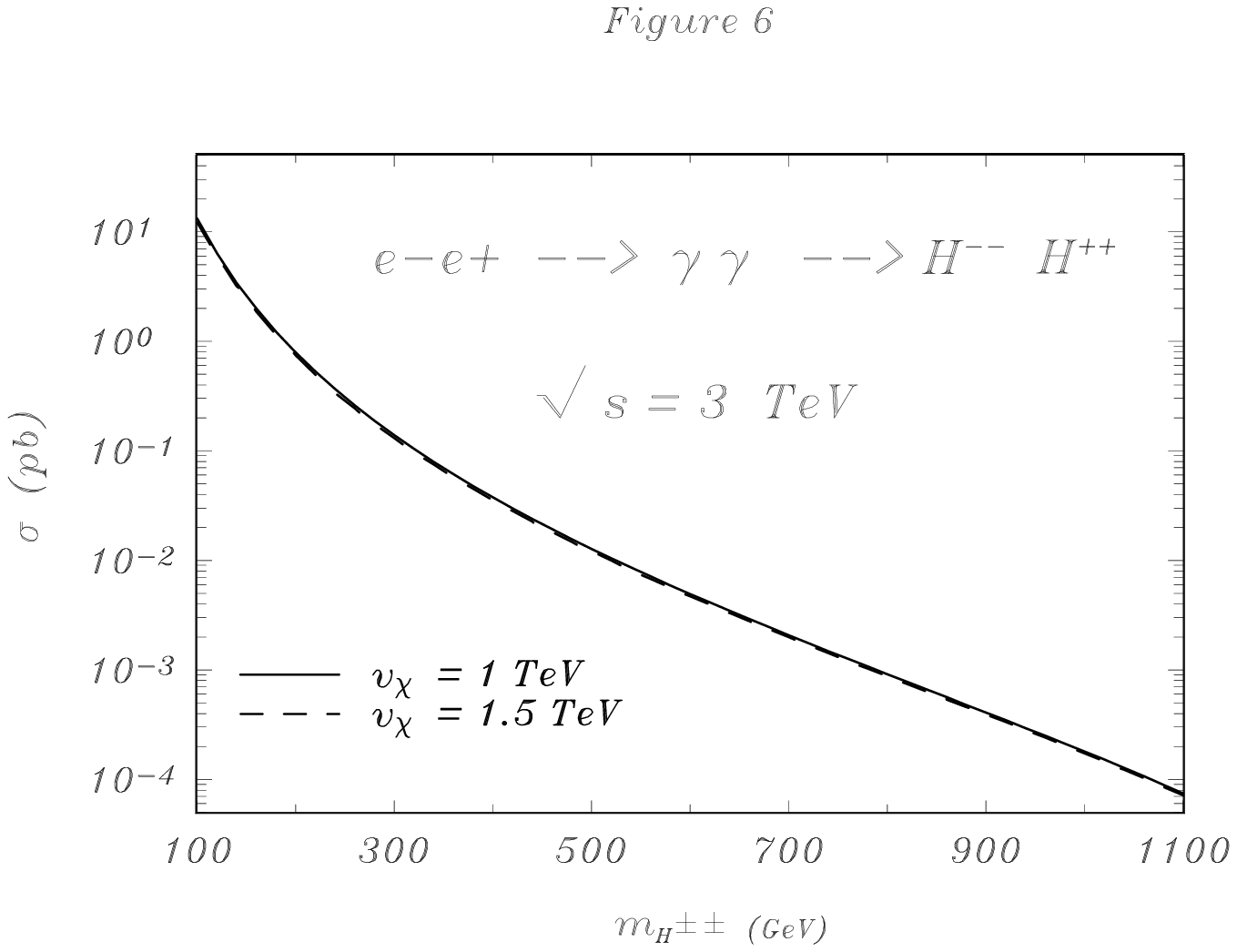}
\caption{\label{fig6} Total cross section for the process $e^{-} e^{+} \rightarrow e^{-} e^{+} \gamma \gamma \rightarrow H^{\pm \pm} H^{\mp \mp}$ as function of $ m_{H^{\pm \pm}}$ for bremstrahlung at $\sqrt{s}=3$ TeV a) $v_{\chi}=1$ TeV (solid line) and b) $v_{\chi}=1.5$ TeV (dashed line).}
\end{figure}

\begin{figure}
\includegraphics [scale=.55]{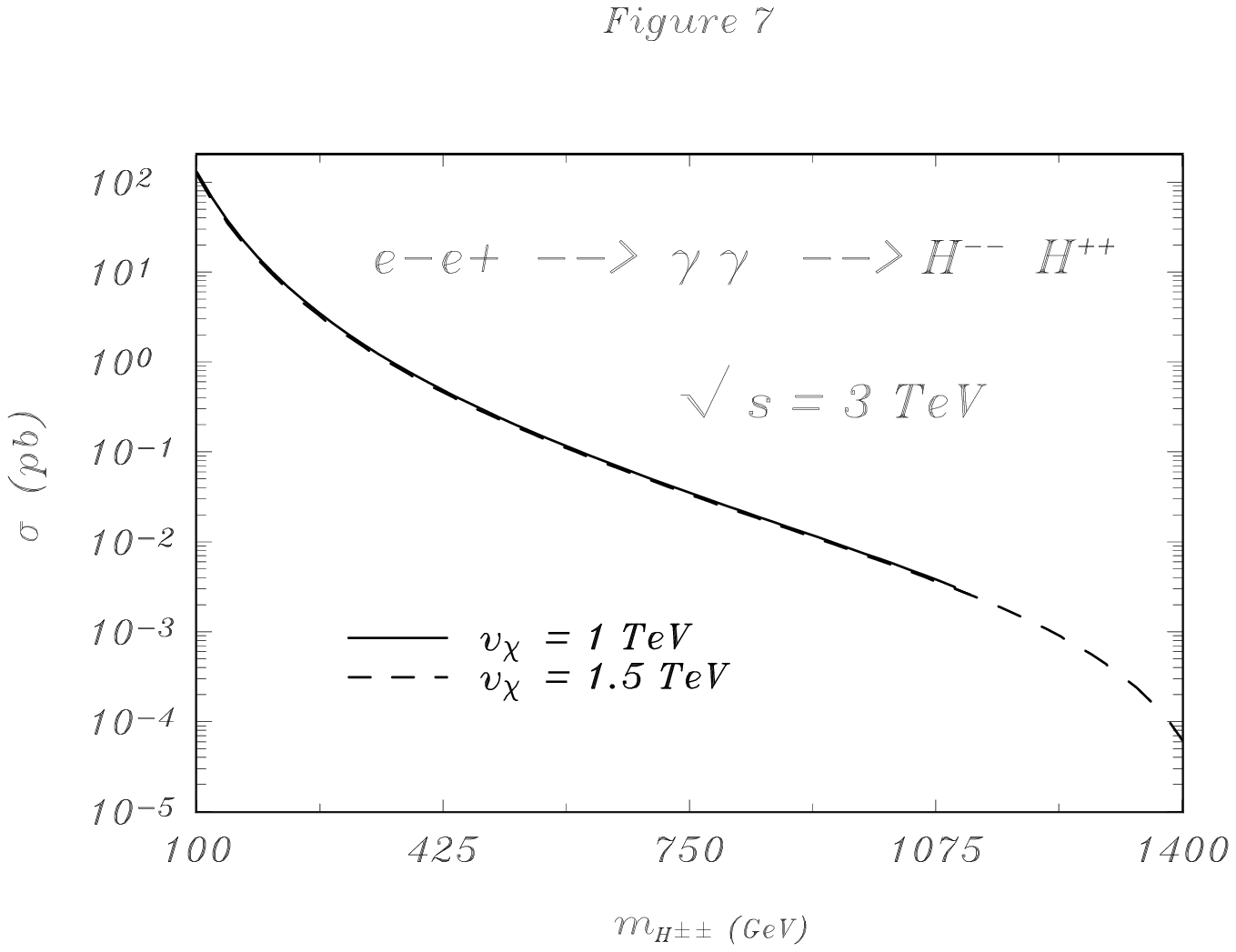}
\caption{\label{fig7} Total cross section for the process $e^{-} e^{+} \rightarrow e^{-} e^{+} \gamma \gamma \rightarrow H^{\pm \pm} H^{\mp \mp}$ as function of $ m_{H^{\pm \pm}}$ for backscattered photons at $\sqrt{s}=3$ TeV a) $v_{\chi}=1$ TeV (solid line) and b) $v_{\chi}=1.5$ TeV (dashed line).}
\end{figure}

\begin{figure}
\includegraphics [scale=.55]{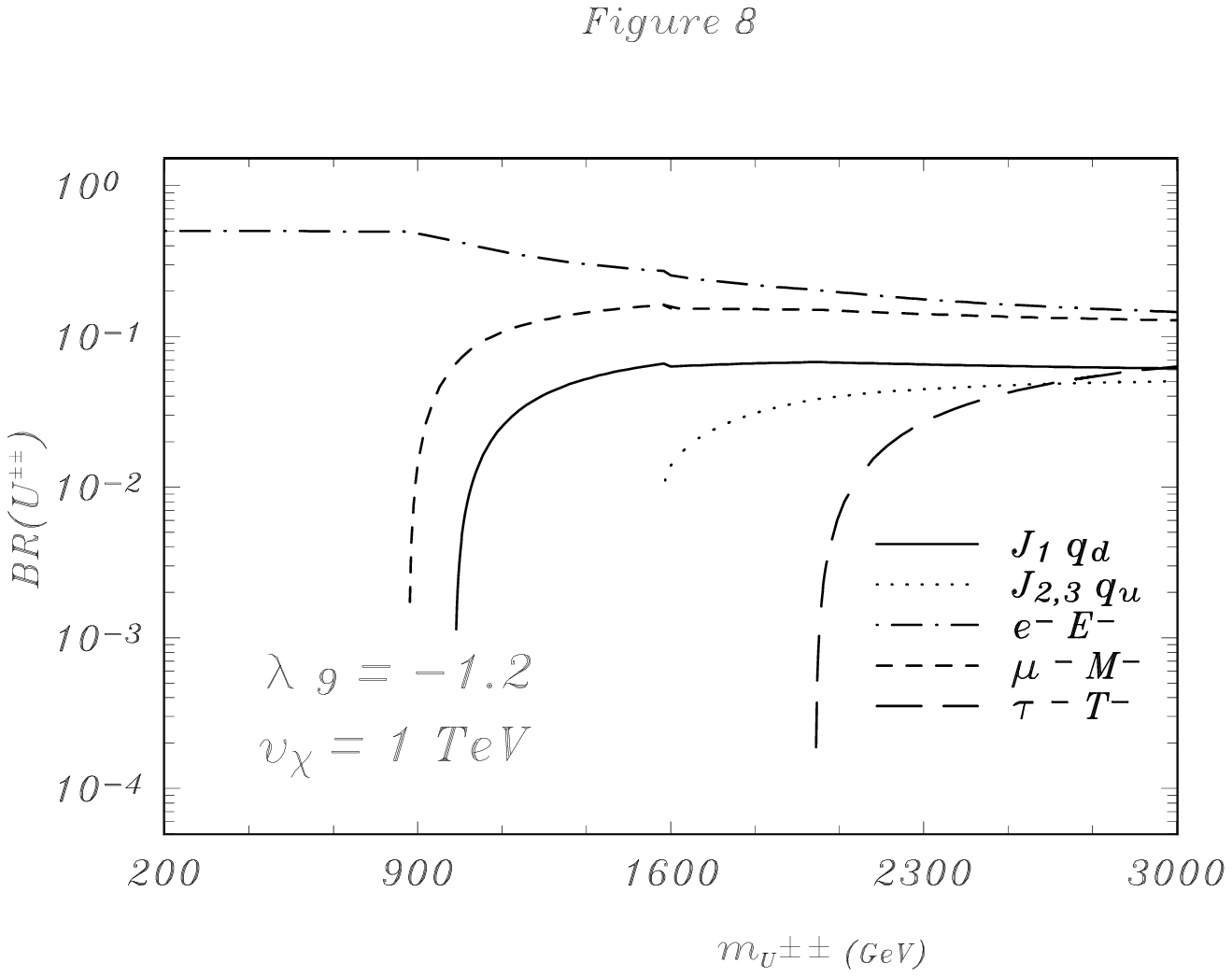}
\caption{\label{fig8} Branching ratios for the doubly charged gauge bosons decays as functions of $m_{H^{\pm \pm}}$ for $\lambda_{9}=- 1.2$, $v_{\chi}=1$ TeV for the quarks ($J_{1} \ q_{d}, J_{2,3} \ q_{u}$) and leptons ($e^{-} \ E^{-}$, $\mu^{-} \  M^{-}$ and $\tau^{-} \ T^{-}$).}
\end{figure}

\begin{figure}
\includegraphics [scale=.55]{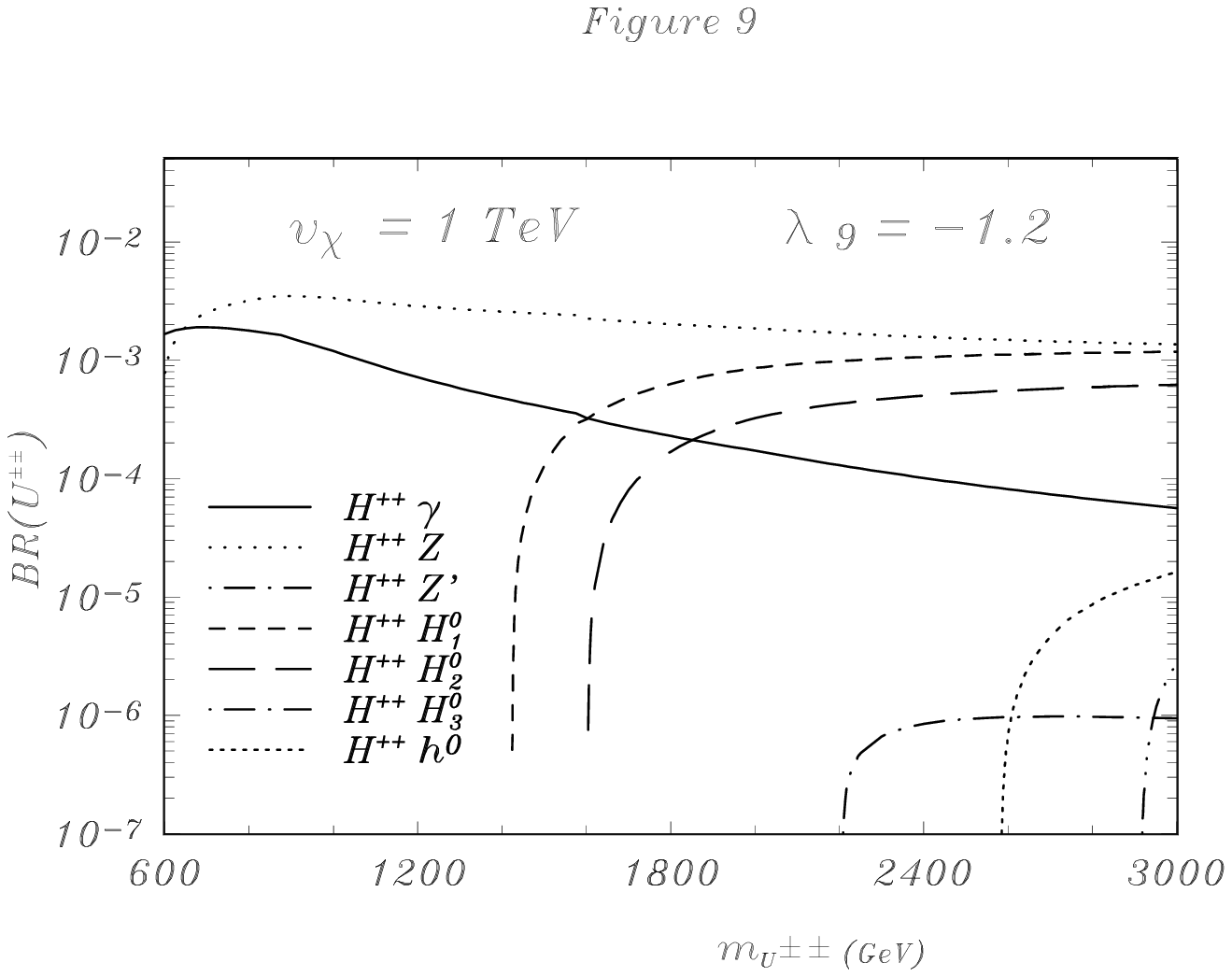}
\caption{\label{fig9} Branching ratios for the doubly charged gauge bosons decays as functions of $m_{H^{\pm \pm}}$ for $\lambda_{9}=- 1.2$, $v_{\chi}=1$ TeV for the doubly charged Higgs and $\gamma$, Z, Z', $H_{1}^{0}$, $H_{2}^{0}$, $H_{3}^{0}$, $h^{0}$.}
\end{figure}
\begin{center}
{\it ILC Collider}\par
\end{center}
Considering that the expected integrated luminosity for the ILC collider will be of order of $3.8  \times 10^5$ pb$^{-1}$/yr, then the statistics give a total of $ \simeq 250$ events per year, if we take the mass of the boson $M_{H^{\pm \pm}}= 500$ GeV and $v_{\chi}=1000$ GeV. Regarding the vacuum  expectation  value $v_{\chi}=1500$ GeV for the same mass it  will give a total of $ \simeq 240$ events per year, in  respect to the backscattered photons (see Fig. 2), the statistics are the following. Taking the mass of the boson $M_{H^{\pm \pm}}= 500$ GeV and $v_{\chi}=1000$ GeV, we will have a total of $\simeq 7.6 \times 10^3$ doubly charged Higgs produced per year.  Regarding the vacuum expectation value $v_{\chi}=1500$ GeV and for the same mass of the boson $H^{\pm \pm}$, it  will give a total of $\simeq 7.2 \times 10^3$ events per year, so we have that for the ILC (bremstrahlung and laser backscattering photons) the number of events is sufficiently appreciable, which we are going now to analyze to detect the signal. \par
Considering that the signal for $H^{\mp \mp}$ are $U^{--} \gamma$ and $U^{++} \gamma$ \cite{cieneto} and taking into account that the branching ratios for these  particles would be $BR(H^{--} \to U^{--} \gamma) = 49.5 \%$ and $BR(H^{++} \to U^{++} \gamma) = 49.5 \%$, see Figs. 3 and 4, for the mass of the Higgs boson $m_{H^{\pm \pm}}= 500$ GeV, $v_{\chi}=1000$ GeV, and that the particles $U^{\mp \mp}$ decay into $e^{-} P^{-}$ and $e^{+} P^{+}$, whose branching ratios for these particles would be $BR(U^{--} \to e^{-} P^{-}) = 50 \% $ and $BR(U^{++} \to e^{+} P^{+}) = 50 \%$, see Figs. 8 and 9, then  we would have approximately 15 events per year for the ILC for 
bremstrahlung photons, regarding the vacuum  expectation value $v_{\chi}=1500$ GeV it  will not give any event because it its restricted by the values of $m_{U^{\pm \pm}}$ which in this case give $m_{U^{\pm \pm}}=691.8$ GeV (see Table I). Considering the backscattered photons, we will have for this signal a total of $\simeq 4.6 \times 10^2$ events per year for the mass of the Higgs boson $m_{H^{\pm \pm}}= 500$ GeV and $v_{\chi}=1000$ GeV.  Regarding the vacuum  expectation value $v_{\chi}=1500$ GeV, it will  not give any event for the same reasons given above.\par
\par
\begin{center}
{\it CLIC Collider}\par
\end{center}
The cross section for the CLIC, which integrated luminosity will be of order of $3 \times 10^6$ pb$^{-1}$/yr, is restricted by the mass of the DCHBs, because for $v_{\chi}=1000(1500)$ and $\lambda_{9}=-1.2$ the acceptable masses are up to $m_{H^{\pm \pm}} \simeq 1107 (1651)$ GeV, see Ref. \cite{cnt2}, taking this into account we obtain for bremstrahlung distribution, see Fig. 6, a total of $\simeq 3.9 \times 10^{4}$ of DCHBs produced per year, if we take the mass of the boson $m_{H^{\pm \pm}}= 500$ GeV and $v_{\chi}=1000$ GeV. Taking the same signal as above, that is, $BR(H^{--} \to U^{--} \gamma) = 49.5 \%$ and $BR(H^{++} \to U^{++} \gamma) = 49.5 \%$, see Figs. 3 and 4, and considering that the particles $U^{\mp \mp}$ decay into $e^{-} P^{-}$ and $e^{+} P^{+}$, whose branching ratios for these particles would be $BR(U^{--} \to e^{-} P^{-}) = 50 \% $ and $BR(U^{++} \to e^{+} P^{+}) = 50 \%$, see Figs. 8 and 9, then  we would have approximately $\simeq 2.3 \times 10^{3}$ events per year for the  CLIC for bremstrahlung photons. Regarding to backscattered photons, see Fig. 7, we would have approximately $7.2 \times 10^5$ of pairs of DCHBs produced per year for the same mass of the Higgs boson $m_{H^{\pm \pm}}= 500$ GeV and $v_{\chi}=1000$ GeV, taking now the same signal as above we will have a total of approximately $4.4 \times 10^4$ events per year for the CLIC collider. It needs to mention here that $P_{a}^{+}$ can decay also as $P_{a}^{+} \rightarrow \ell^{-} \ell^{+} P_{b}^{+}$, where the index $a$ denotes the flavor, therefore we will have eight charged leptons, see Ref. \cite{cieton3}. 

We still mention here that there exist an asymmetry between $H^{--}$ and $H^{++}$ to respect to decay rates we have not considered, such as $H^{\pm \pm} \to e^{\pm} P^{\pm}$, where the rate of $H^{- -}  \to e^{-} P^{-}$ is largeer than the rate of $H^{+ +} \to e^{+} P^{+}$ ( Fig. 3), which indicates that in this 3-3-1 model the $H^{- -}$ generate matter in a large rate compared to rate of the  $H^{++}$, that generates antimatter. In this way symmetry does not exist between the two  DCHBs to respect to decay rates. We can do an indirect measurement of this fact measuring the  other rates such as $H^{\pm \pm} \to U^{\pm \pm} \gamma \to e^{\pm} P^{\pm}$, which was calculated here.       

So the $\gamma \gamma$ collisions can  be also a plentiful source of DCHBs \cite{chak,sura}. In relation to the signal, $H^{\pm \pm} \to U^{\pm \pm} \gamma$ and $U^{\pm \pm} \to e^{\pm} P^{\pm}$, we conclude  that it is a very striking and important signal, the DCHBs will deposit six times the ionization energy than the characteristic single-charged particle, that is, if we see this signal we will not only be seeing the DCHBs but also the doubly charged gauge bosons and heavy leptons. \par
It is to notice here that there are no backgrounds to this signal. The discovery of a pair of DCHBs will be without any doubt of great importance for the physics beyond the SM,  because of the confirmation of the Higgs triplet representation and indirect verification that there is asymmetry in decay rates between matter  and antimatter. \par

In summary, through this work, we have shown that in the context of the 3-3-1 model the signatures for DCHBs can be significant for $\gamma \gamma$ collisions obtained  by bremstrahlung and backward Compton scattering of laser. Our study indicates the possibility of obtaining a clear signal of these new particles with a satisfactory number of events.


\end{document}